\documentclass[aps,pra,showpacs,twocolumn,amsmath,amssymb,groupedaddress]{revtex4-1}
\usepackage{graphicx}
\begin{document}

\title{Interface solitons in thermal nonlinear media}
% Force line breaks with \\

\author{Xuekai Ma, Zhenjun Yang, Daquan Lu, and Wei Hu}

\email[Corresponding author's email address: ]{huwei@scnu.edu.cn}
%\thanks{}

\affiliation{Laboratory of Photonic Information Technology, South China Normal University, Guangzhou 510631, P. R. China}

\date{\today}% It is always \today, today,

\begin{abstract}
We demonstrate the existence of fundamental and dipole interface solitons in one-dimensional thermal nonlinear media with a step in linear refractive index.  Fundamental interface solitons are found to be always stable and the stability of dipole interface solitons depends on the difference in linear refractive index. The mass center of interface solitons always locates in the side with higher refractive index. Two intensity peaks of dipole interface solitons are unequal except some specific conditions, which is different from their counterparts in uniform thermal nonlinear media.
\end{abstract}

\pacs{42.65.Tg, 42.65.Jx}

\maketitle

Nonlocal solitons have been found in many physical systems, such as photorefractive crystals~\cite{Mitchell1998-PRL}, nematic liquid crystals~\cite{Conti2003-PRL,Conti2004-PRL}, lead glasses~\cite{Rotschild2005-PRL,Rotschild-NaturePhysics-2006}, atomic vapors~\cite{Skupin2007-PRL}, and Bose-Einstein condensates~\cite{Pedri2005-PRL,Tikhonenkov2008-PRL}, etc. In nonlocal nonlinear media, various types of solitons have been studied theoretically and experimentally, such as vortex solitons~\cite{Kartashov2007-OE1}, multipole solitons~\cite{Rotschild2006-OL,Izdebskaya2011-OL}, Laguerre and Hermite soliton clusters~\cite{Buccoliero2007-PRL}, Ince-Gaussian solitons~\cite{Deng2007-OL}. For nonlocal solitons, there are many interesting properties, for instance,  large phase shift~\cite{Guo2004-PRE}, attraction between two dark solitons~\cite{Dreischuh2006-PRL,Nikolov2004-OL}, self-induced fractional Fourier transform~\cite{Lu2008-PRA}, etc.

Surface waves are localized waves residing at the interface between two media with different optical properties. They have been used to study the surface properties of media in physics, chemistry, and biology. In the presence of nonlinearity, some kinds of surface solitons have been found theoretically and experimentally. Recently, nonlocal surface solitons have been studied both theoretically and experimentally~\cite{Alfassi-2007-PRL,Kartashov-2009-OL,Ye-2008-PRA,Alfassi-2009-PRA,Kartashov-2006-OL}. It is proved that the nonlocality can support various types of surface solitons, such as multipole surface solitons~\cite{Kartashov-2009-OL,Ye-2008-PRA,Kartashov-2006-OL}, vortex surface solitons~\cite{Ye-2008-PRA}, and incoherent surface solitons~\cite{Alfassi-2009-PRA}. Surface dipoles, vortices, and bound states of vortex solitons are found to be stable at two dimensional interfaces~\cite{Ye-2008-PRA}. In one dimension case, multipole surface solitons are stable when the number of poles is less than three, whereas the higher-order solitons can be stable in layered thermal media~\cite{Kartashov-2009-OL}. When the interface is formed by two nonlocal nonlinear media, both fundamental and dipole interface solitons are found to be stable in the presence of the optical lattice, but dipole interface solitons do not exist in  uniform latticeless media~\cite{Kartashov-2006-OL}. Here we study the surface waves at the interface formed by two thermal nonlinear media, in which the boundary force effect~\cite{Alfassi-2007-OL,Shou-2009-OL} can support the existence of dipole interface solitons.

In this paper, we demonstrate the existence of interface solitons in the thermal nonlinear media which own two different linear refractive indices. Stability analysis of interface solitons are carried out. It is found that fundamental interface solitons are always stable and the stability of dipole interface solitons depends on the refractive index difference between two media. Two intensity peaks of dipole interface solitons are unequal except some specific conditions, which is different from their counterparts in uniform thermal media.

We consider a (1+1)dimensional thermal sample occupying the region $-L\leq x\leq L$. The sample is separated into two parts at the center ($x=0$). Two boundaries ($x=\pm L$) and the interface are thermally conductive. All parameters of the two parts, such as thermal conductivity coefficient, absorption coefficient and thermal coefficient, are the same except the linear refractive index. The propagation of a transverse-electric (TE)  polarized laser beam is governed by the dimensionless nonlocal nonlinear Schr\"{o}dinger equation\\
(i) in the left, i.e. $-L\leq x\leq0$
\begin{equation}\label{1}
  i\frac{\partial q}{\partial
z}+\frac{1}{2}\frac{\partial^2q}{\partial x^2}+nq=0,\,\,\,\,\,\,\,
  \frac{\partial^2n}{\partial x^2}=-|q|^2;
\end{equation}
(ii) in the right, i.e. $0\leq x\leq L$
\begin{equation}\label{2}
  i\frac{\partial q}{\partial
z}+\frac{1}{2}\frac{\partial^2q}{\partial x^2}+nq-n_d q=0,\,\,\,\,
  \frac{\partial^2n}{\partial x^2}=-|q|^2,
\end{equation}
where $x$ and $z$ stand for the transverse and longitudinal coordinates scaled to the beam width and the diffraction length, $q$ is the complex amplitude of optical field, $n$ is the nonlinear refractive index change, and $n_d>0$ is proportional to the linear refractive index difference between two media.

The boundary conditions can be described as $q(\pm L)=0$ when the laser beam is narrow and far from two boundaries. The continuity conditions at the interface ($x=0$)  depend on the polarization. For the TE polarized wave, the continuity conditions for the transverse field are $q(-0)=q(+0)$ and  $\partial q/\partial x|_{x=-0}=\partial q/\partial x|_{x=+0}$ \cite{Alfassi-2007-PRL}. For the nonlinear refractive index, we have the boundary conditions  $n(\pm L)=0$ because two boundaries are thermally stabilized by means of external heat sinks \cite{Alfassi-2007-PRL,Kartashov-2009-OL}.  Since the interface is also thermally conductive, we have the continuity relation $n(-0)=n(+0)$. From the continuity of $q$ and the nonlinear equations in Eqs. (\ref{1}) and (\ref{2}), the derivative of nonlinear refractive index is also continuous.

\begin{figure}[htb]
\centerline{\includegraphics[width=8.6cm]{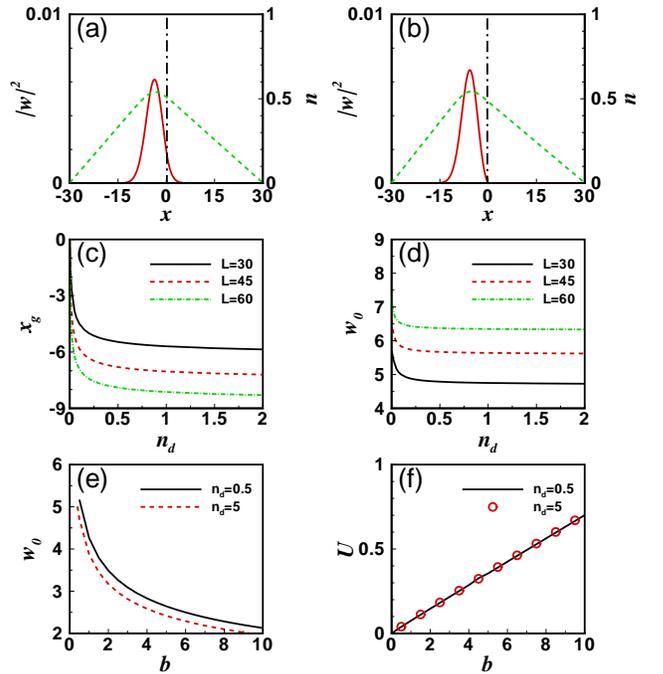}}
  \caption{(Color online) Profiles of fundamental interface solitons at (a) $n_d=0.05$, $b=0.5$ and (b) $n_d=5$, $b=0.5$. Solid red lines stand for optical intensity while dashed green lines stand for nonlinear refractive index. Dash-dotted line indicates the interface and $L=30$. (c) Soliton mass center versus $n_d$ at $b=0.5$ for different sample sizes. (d) Beam width versus $n_d$ at $b=0.5$ for different sample sizes. (e) Beam width versus propagation constant at different $n_d$ for $L=30$. (f) Energy flows $U$ versus propagation constant at different $n_d$ for $L=30$. }\label{F1}
\end{figure}

\begin{figure}[htb]
\centerline{\includegraphics[width=8.6cm]{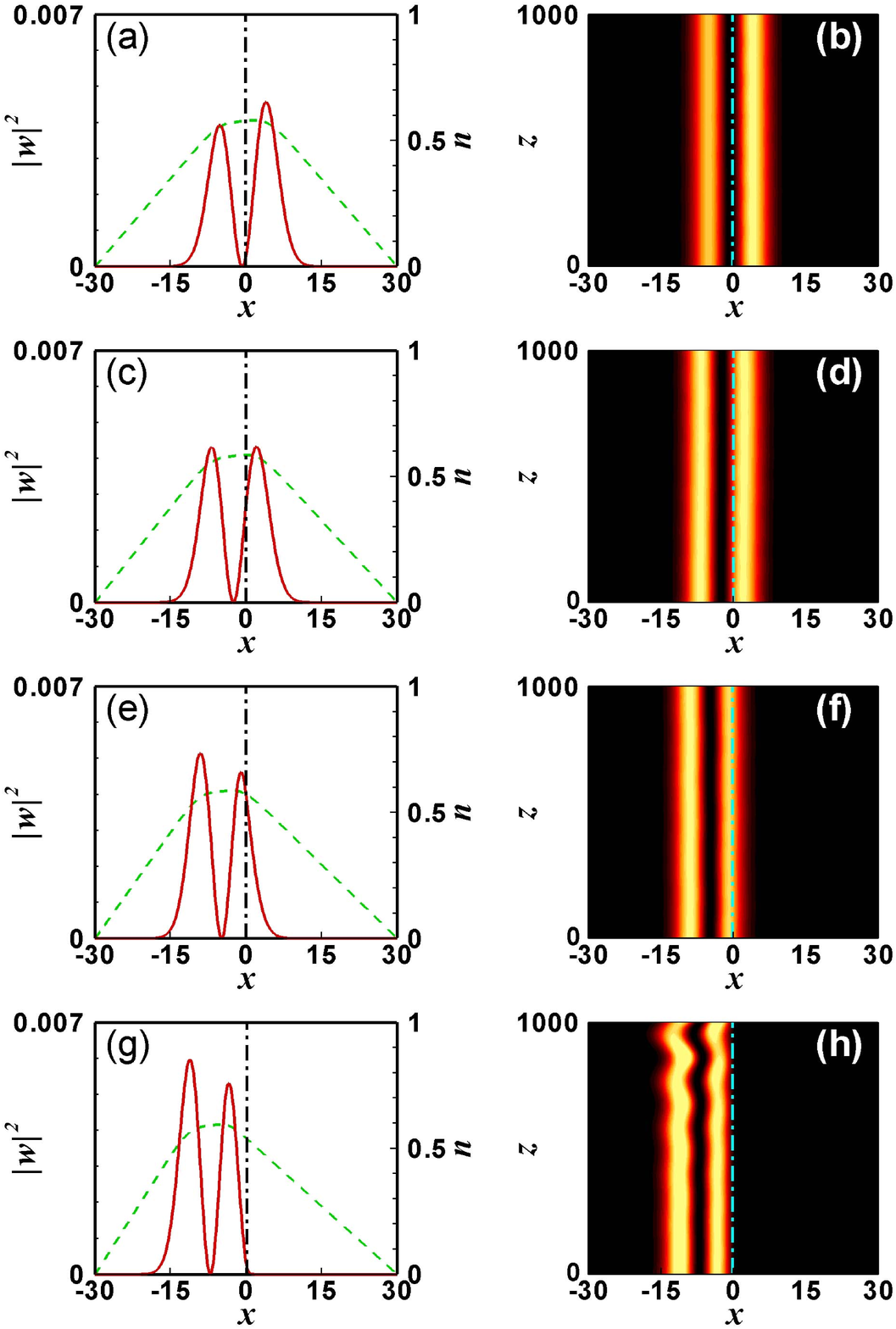}}
 \caption{(Color online) Profiles of dipole interface solitons at (a)$n_d=0.02$, (c) $n_d=0.029$, (e) $n_d=0.05$, and (g) $n_d=1$. Solid red lines stand for optical intensity while dashed green lines stand for nonlinear refractive index. (b), (d), (f) and (h) denote the propagations of dipole interface solitons corresponding to (a), (c), (e) and (f), respectively. Note the dipole interface soliton at $n_d=1$ in (g) and (h) is unstable. In all cases $b=0.5$, $L=30$. }\label{F2}
\end{figure}

We search for the soliton solutions for Eqs.~(\ref{1}) and (\ref{2}) in the form $q(x,z)=w(x)\exp(ibz)$, where $w(x)$ is a real function, $b$ is the propagation constant. An iterative method is used to get numerical solutions for different $n_d$ and $b$.  The results for fundamental and dipole interface solitons are shown in Figs.~\ref{F1} and \ref{F2}, respectively.

In order to elucidate the stability of interface solitons, we search for the perturbed solutions of Eqs.~(\ref{1}) and (\ref{2}) in the form $q=(w+u+iv)\exp(ibz)$, where $u(x,z)$ and $v(x,z)$ are the real and the imaginary parts of small perturbations. The perturbation can grow with a complex rate $\sigma$ upon propagation. Substituting the perturbed soliton solution into Eqs. (\ref{1}) and (\ref{2}), and using the stationary soliton solution $w(x)$, one can get the linear eigenvalue problem around stationary solution $w(x)$, \begin{equation}\label{3}
 \left.\begin{aligned}
  \sigma u&=-\frac{1}{2}\frac{d^2v}{dx^2}+bv-nv,  \\
  \sigma v&=\frac{1}{2}\frac{d^2u}{dx^2}-bu+nu+w\Delta n,  \\
\end{aligned}\right\}
\,\,\,\, \text{($-L\leq x\leq0$),}
\end{equation}
and
\begin{equation}\label{4}
 \left.\begin{aligned}
  \sigma u&=-\frac{1}{2}\frac{d^2v}{dx^2}+bv-nv+n_dv,  \\
  \sigma v&=\frac{1}{2}\frac{d^2u}{dx^2}-bu+nu-n_du+w\Delta n,  \\
\end{aligned}\right\}
\,\,\,\, \text{($0\leq x\leq L$),}
\end{equation}
where $\Delta n=-2\int_{-L}^{L}G(x,x')w(x')u(x')dx'$ is the refractive index perturbation, the response function $G(x,x')=(x+L)(x'-L)/(2L)$ for $x\leq x'$ and $G(x,x')=(x'+L)(x-L)/(2L)$ for $x\geq x'$, $\sigma_r$ (real part of $\sigma$) represents the instability growth rate.

The eigenvalue problem of Eqs. (\ref{3}) and (\ref{4}) has been solved numerically. We find that the fundamental interface solitons are always stable in their whole domain for both small and large $n_d$. For comparison, we know that an optical beam can form a stable fundamental soliton in bulk thermal nonlocal media~\cite{Dong-2010-PRA,Xu-2005-OL}, and fundamental surface solitons are also stable in thermal nonlocal media~\cite{Kartashov-2009-OL}.

First, we discuss fundamental interface solitons as shown in Fig. \ref{F1}. In our model, when $n_d$ approaches zero, the interface soliton reduces to a bulk soliton. As $n_d$ increases, the stable interface soliton shifts itself to the higher index part. The soliton mass center, defined as $x_g=\int_{-\infty}^{\infty}x|q|^2dx/\int_{-\infty}^{\infty}|q|^2dx$, will always locate in the left part where the linear refractive index is higher. For example, for $n_d=0.05$ as shown in Fig. \ref{F1}(a), $x_g=-3.8$ and most energy of the soliton resides in the left part, while a little energy resides in the right. $|x_g|$ increases monotonically as $n_d$ increases, as shown in Fig.~\ref{F1}(c). When $n_d>1$, almost all energy of the soliton resides in the left part [Fig. \ref{F1}(b)]. The intensity profiles for larger $n_d$ are similar to those of surface solitons~\cite{Alfassi-2007-PRL,Kartashov-2009-OL}, whereas the nonlinear index changes are different. For  surface solitons, there is a step in the nonlinear index at the interface ~\cite{Alfassi-2007-PRL,Kartashov-2009-OL}, but the nonlinear index distribution is continuous at the interface in our model.

For much larger $n_d$, $x_g$ and $w_0$ approach to certain values, i.e. $x_g\rightarrow-6$ and  $w_0\rightarrow4.7$ for  $b=0.5$ and $L=30$ as shown in Figs.~\ref{F1}(c) and \ref{F1}(d).  Because the boundary force effect~\cite{Alfassi-2007-OL,Shou-2009-OL} always pushes the soliton to the center of the sample, the interface soliton can not shift away from the interface.  It is obvious that the asymptotic value of $x_g$ is proportional to the asymptotic beam width when $n_d$ approaches to infinity. The asymptotic value of $|x_g|$  decreases as the propagation constant $b$ increases [same as the beam width shown in Fig.  \ref{F1}(e)], but increases as the sample size $L$ increases [Fig. \ref{F1}(c)].  As shown in Fig. \ref{F1}(f), the energy flow, defined as $U=\int_{-\infty}^{\infty}|q|^2dx$, is a linearly growing function of $b$ for different $n_d$. It is worthy to note that the change of $n_d$ does not influence the energy flow of the fundamental interface solitons because the peak intensity increases while the beam width decreases.

\begin{figure}[htb]
\centerline{\includegraphics[width=8.6cm]{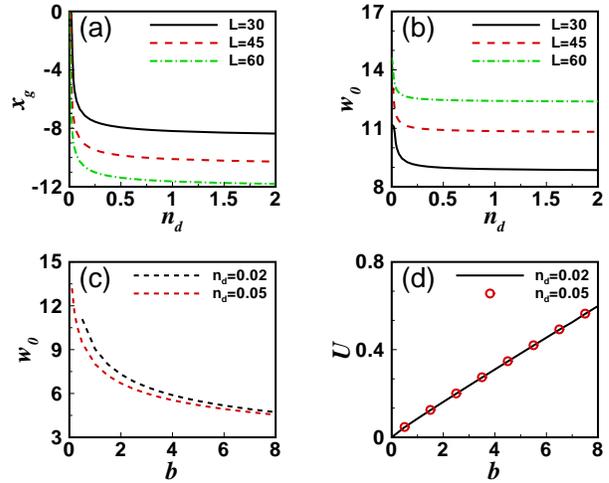}} \caption{(Color
online)(a) Dipole mass center versus $n_d$ at $b=0.5$ for different sample sizes. (b) Beam width versus $n_d$ at $b=0.5$ for different sample sizes. (c) Beam width versus propagation constant at different $n_d$ for $L=30$. (d) Energy flows $U$ versus propagation constant at different $n_d$ for $L=30$.}\label{newF3}
\end{figure}

The results of dipole interface solitons are shown in Figs. \ref{F2} and \ref{newF3}.  The change of the two intensity peaks of interface dipoles is interesting. When $n_d$ approaches to zero, similar to a bulk dipole soliton, the two intensity peaks are equal. As $n_d$ increases, the two peaks increase at different rates and the right peak becomes higher than the left one [Figs. \ref{F2}(a) and \ref{F2}(b)], whereas the mass center moves toward the left [Fig. \ref{newF3}(a)]. Then the left peak increases more quickly, and the two peaks become equal again when $n_d=0.029$ [Figs. \ref{F2}(c) and \ref{F2}(d)]. Increasing $n_d$ sequentially, the left peak becomes higher than the right one and a significant part of the right peak resides in the left part [Figs. \ref{F2}(e) and \ref{F2}(f)]. When $n_d>0.5$, such as $n_d=1$ shown in Figs. \ref{F2}(g) and \ref{F2}(h), almost all energy moves into the left part and the intensity profiles are similar to those of dipole surface solitons, however, their nonlinear refractive index distributions are different~\cite{Kartashov-2009-OL}. Similar to fundamental interface solitons, the soliton center moves slowly toward the left and the beam width decreases monotonically as $n_d$ increases [Figs. \ref{newF3}(a) and \ref{newF3}(b)]. In Fig.~\ref{newF3}(c) and \ref{newF3}(d), the beam width decreases monotonically with increasing $b$, whiereas the energy flows of dipole interface solitons increase monotonically with increasing $b$. The energy flows for different $n_d$ are almost equal for the same reason mentioned in fundamental interface solitons.

\begin{figure}[htb]
\centerline{\includegraphics[width=8.6cm]{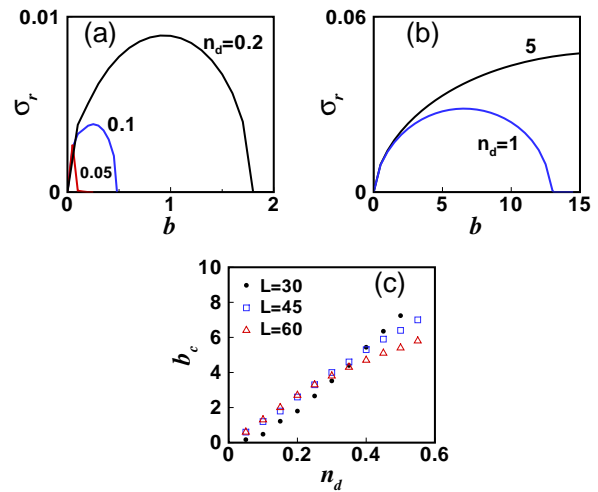}} \caption{(Color
online) Real part of the perturbation growth rate for dipole interface
 solitons at (a) $n_d=0.2, 0.1, 0.05$, and (b) $n_d=1, 5$. In those
cases $L=30$. (c) Critical value of propagation constant versus $n_d$ for different sample sizes.  }\label{F3}
\end{figure}

An important result of this paper is that the stability of dipole interface solitons depends on the index difference $n_d$. For comparison, dipole surface solitons in thermal nonlocal media are stable~\cite{Kartashov-2009-OL}, but dipole interface solitons in nonlocal nonlinear media with a finite range of nonlocality are unstable~\cite{Kartashov-2006-OL}. Figure \ref{F3} presents the results of the stability analysis for dipole interface solitons. For a given $n_d$, there exists a critical propagation constant $b_c$, and dipoles are stable when $b>b_c$. For $n_d=0.2$ and $0.1$, the stable regions are $b>1.8$ and  $b>0.5$, respectively [Fig.~\ref{F3}(a)]. The relation between critical propagation constant $b_c$ and $n_d$ is shown in Fig.~\ref{F3}(c) for different sample sizes $L$.  If $n_d$ is very small, the dipole interface solitons are stable almost in their whole domain, which consists with the dipole solitons in bulk nonlocal media~\cite{Dong-2010-PRA,Xu-2005-OL}.

In Ref.~\cite{Kartashov-2006-OL}, lattices are found to be necessary for the dipole solitons at the interface of two nonlocal nonlinear media with a finite range of nonlocality. Here the thermal nonlinearity has an infinite range of nonlocality and the boundaries are essential~\cite{Alfassi-2007-PRL}. The boundary force effect, which vanishes in the nonlocal nonlinear media with a finite range of nonlocality, can support the stability of dipole interface solitons, whatever small or large $n_d$. From Fig.~\ref{F3}(c), $b_c$ varies slightly when the sample size is doubled. It implies that the boundary force effect do not decrease significantly when the boundaries move far away from the solitons. Due to the infinite range of nonlocality, dipole interface solitons can exist in very large samples.

\begin{figure}[htb]
\centerline{\includegraphics[width=8.6cm]{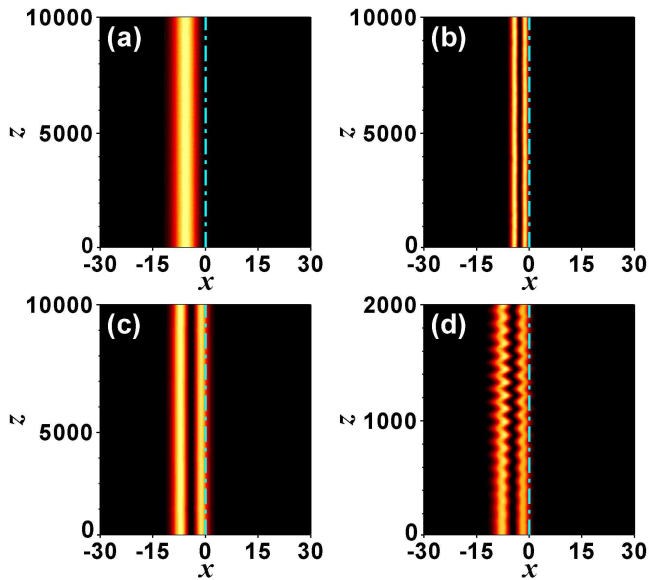}} \caption{(Color
online) (a) Propagation of a perturbed interface fundamental soliton
at $n_d=5$, $b=0.5$. Propagations of perturbed dipole interface
solitons at (b) $n_d=1$, $b=16$, (c) $n_d=0.1$, $b=1.5$, and (d)
$n_d=0.2$, $b=1.5$.  In all
cases $L=30$.}\label{F4}
\end{figure}

To confirm the results of the linear stability analysis, we simulate the soliton propagations based on Eqs. (\ref{1}) and (\ref{2}) with the input condition $q(x,z=0)=w(x)[1+\rho(x)]$, where $w(x)$ is the profile of the stationary wave and $\rho(x)$ is a random function which stands for the input noise with the variance $\delta_{noise}^2=0.01$. Figure \ref{F4} presents the propagations of fundamental and dipole interface solitons. As expected, the fundamental and dipole interface solitons at the stable region predicted by the linear stability analysis survive over long propagation distance in the presence of the input noise [Figs.~\ref{F4}(a)-(c)]. Figure \ref{F4}(d) presents a dipole soliton in the instability region, it experiences oscillatory instability after propagating over a certain distance.

To conclude, we have presented the study of interface solitons in thermal nonlinear media. This type of solitons shows some unique properties which differ from those of bulk solitons and surface solitons. The linear stability analysis shows that fundamental interface solitons are always stable and the stability of dipole interface solitons depends on the difference in the linear refractive index.  We consider that the boundary force effect plays an important role in the stability of dipoles. As the index difference $n_d$ approaches to zero, our interface solitons reduce to the corresponding bulk solitons which have been proved to be stable. For large $n_d$, the interface solitons are similar to the surface solitons in intensity profiles, but different in nonlinear refractive indices.

The authors gratefully acknowledge useful comments from the referee. This research was supported by the National Natural Science Foundation of China (Grant Nos. 10804033 and 10674050), the Program for Innovative Research Team of Higher Education in Guangdong (Grant No. 06CXTD005), and the Specialized Research Fund for the Doctoral Program of Higher Education (Grant No. 200805740002).


\begin{thebibliography}{99}


\bibitem{Mitchell1998-PRL}
M. Mitchell, M. Segev, and D. N. Christodoulides,
%``Observation of multihump multimode solitons,''
\prl  {\bf 80,} 4657-4660 (1998).

\bibitem{Conti2003-PRL}
C. Conti, M. Peccianti, and G. Assanto,
%``Route to nonlocality and observation of accessible solitons,''
\prl  {\bf 91,} 073901 (2003).

\bibitem{Conti2004-PRL}
C. Conti, M. Peccianti, and G. Assanto,
%``Observation of optical spatial solitons in a highly nonlocal medium,''
\prl  {\bf 92,} 113902 (2004).


\bibitem{Rotschild2005-PRL}
C. Rotschild, O. Cohen, O. Manela, M. Segev, and T. Carmon,
%``Solitons in nonlinear media with an infinite range of nonlocality: first
%observation of coherent elliptic solitons and of vortex-ring solitons,''
\prl  {\bf 95,} 213904 (2005).

\bibitem{Rotschild-NaturePhysics-2006} C. Rotschild, B. Alfassi, O. Cohen, and M. Segev,
% "Long-range interactions between optical solitons",
Nature Phys. {\bf 2}, 769-774 (2006).

\bibitem{Skupin2007-PRL}
S. Skupin, M. Saffman, and W. Krolikowski,
%``Nonlocal stabilization of nonlinear beams in a self-focusing atomic vapor,''
\prl  {\bf 98}, 263902 (2007)

\bibitem{Pedri2005-PRL}
P. Pedri and L. Santos,
% ``Two-dimensional bright solitons in dipolar Bose-Einstein condensates,''
\prl {\bf 95}, 200404 (2005)

\bibitem{Tikhonenkov2008-PRL}
I. Tikhonenkov, B. A. Malomed, and A. Vardi,
%``Anisotropic solitons in dipolar Bose-Einstein condensates,''
\prl {\bf 100}, 090406 (2008)

\bibitem{Kartashov2007-OE1}
Y. V. Kartashov, V. A. Vysloukh, and L. Torner,
%``Stability of vortex solitons in thermal nonlinear media with cylindrical symmetry,
Opt. Express {\bf 15}, 9378-9384 (2007),

\bibitem{Rotschild2006-OL}
C. Rotschild, M. Segev, Z. Xu, Y. V. Kartashov, L. Torner, and O. Cohen,
%``Two-dimensional multipole solitons in nonlocal nonlinear media,''
\ol {\bf 31}, 3312-3314 (2006).

\bibitem{Izdebskaya2011-OL}
Y. V. Izdebskaya, A. S. Desyatnikov, G. Assanto, and Y. S. Kivshar,
%``Multimode nematicon waveguides,''
\ol {\bf 36}, 184-186 (2011).

\bibitem{Buccoliero2007-PRL}
D. Buccoliero, A. S. Desyatnikov, W. Krolikowski, and Y. S. Kivshar,
%``Laguerre and Hermite soliton clusters in nonlocal nonlinear media,''
\prl {\bf 98}, 053901 (2007).

\bibitem{Deng2007-OL}
D. Deng and Q. Guo,
%``Ince-Gaussian solitons in strongly nonlocal nonlinear media,''
\ol {\bf 32}, 3206-3208 (2007).

\bibitem{Guo2004-PRE}
Q. Guo, B. Luo, F. H. Yi, S. Chi, and Y. Q. Xie,
%``Large phase shift of nonlocal optical spatial solitons,''
\pre  {\bf 69,} 016602 (2004).


\bibitem{Dreischuh2006-PRL}
A. Dreischuh, D. N. Neshev, D. E. Petersen, O. Bang, and W. Krolikowski,
%``Observation of attraction between dark solitons,''
\prl {\bf 96,} 043901 (2006).

\bibitem{Nikolov2004-OL}
N. I. Nikolov, D. Neshev, W. Krolikowski, O. Bang, J. J. Rasmussen,
and P. L. Christiansen,
%``Attraction of nonlocal dark optical solitons,''
\ol {\bf 29,} 286-288 (2004).

\bibitem{Lu2008-PRA}
D. Q. Lu, W. Hu, Y. J. Zheng, Y. B. Liang, L. G. Cao, S. Lan, and Q. Guo,
% ``Self-induced fractional Fourier transform and revivable higher-order spatial
% solitons in strongly nonlocal nonlinear media,''
\pra {\bf 78,} 043815 (2008).

\bibitem{Alfassi-2007-PRL}
B. Alfassi, C. Rotschild, O. Manela, M. Segev, and D. N. Christodoulides,
%``Nonlocal Surface-Wave Solitons,''
\prl {\bf98}, 213901 (2007).

\bibitem{Ye-2008-PRA}
F. Ye, Y. V. Kartashov, and L. Torner,
%``Nonlocal surface dipoles and vortices,''
\pra {\bf77}, 033829 (2008).

\bibitem{Kartashov-2009-OL}
Y. V. Kartashov, V. A. Vysloukh, and L. Torner,
%``Multipole surface solitons in thermal media,''
\ol {\bf34}, 283-285 (2009).

\bibitem{Kartashov-2006-OL}
Y. V. Kartashov, L. Torner, and V. A. Vysloukh,
%"Lattice-supported surface solitons in nonlocal nonlinear media"
\ol {\bf 31}, 2595-2597 (2006).

\bibitem{Alfassi-2009-PRA}
B. Alfassi, C. Rotschild, and M. Segev,
%``Incoherent surface solitons in effectively instantaneous nonlocal nonlinear media,''
\pra{\bf 80}, 041808 (2009).

\bibitem{Alfassi-2007-OL} B. Alfassi, C. Rotschild, O. Manela, M. Segev, and D. N. Christodoulides
%"Boundary force effects exerted on solitons in highly nonlocal nonlinear media"
\ol {\bf 32}, 154-156 (2007).

\bibitem{Shou-2009-OL}Q. Shou, Y. B. Liang, Q. Jiang, Y. J. Zheng, S. Lan, W. Hu, and Q. Guo,
%¡°Boundary force exerted on spatial solitons in cylindrical strongly nonlocal media,¡±
\ol {\bf 34}, 3523-3525 (2009).


\bibitem{Xu-2005-OL}
Z. Xu, Y. V. Kartashov, and L. Torner,
% ``Upper threshold for stability of multipole-mode solitons in nonlocal nonlinear media,''
\ol {\bf30}, 3171-3173 (2005).


\bibitem{Dong-2010-PRA}
L. Dong and F. Ye,
%``Stability of multipole-mode solitons in thermal nonlinear media,''
\pra {\bf81}, 013815 (2010).


\end{thebibliography}
\end{document}